\documentclass[11pt,showpacs,preprintnumbers,amsmath,amssymb,prd,nofootinbib,superscriptaddress]{revtex4-2}

\usepackage{dcolumn}
\usepackage{bm}
\usepackage{ifpdf}
\usepackage{hyperref}
\usepackage{bm}
\usepackage{xcolor,color,graphicx,graphics,physics}
\usepackage[spanish,english]{babel}
\usepackage[latin1]{inputenc}
\usepackage[OT1]{fontenc}
\usepackage{latexsym,amssymb,amsmath,amsfonts, slashed}
\usepackage{makeidx}
\usepackage{epsfig,subfigure}
\usepackage{natbib}
\usepackage{epstopdf}
\usepackage{mathrsfs}
\usepackage{hyperref}
\hypersetup{colorlinks=true, linkcolor=blue, citecolor=blue, urlcolor=blue}
\usepackage{enumerate}

\usepackage{fixmath}

\everymath{\displaystyle}
\usepackage{graphicx}

\usepackage[T1]{fontenc}
\usepackage{amsmath}
\usepackage{amssymb}
\usepackage{graphicx}
\usepackage{xcolor}

\usepackage{url}

\newcommand{\bea}{\begin{eqnarray}}
\newcommand{\eea}{\end{eqnarray}}

\newcommand{\orcid}[1]{\href{https://orcid.org/#1}{\includegraphics[width=10pt]{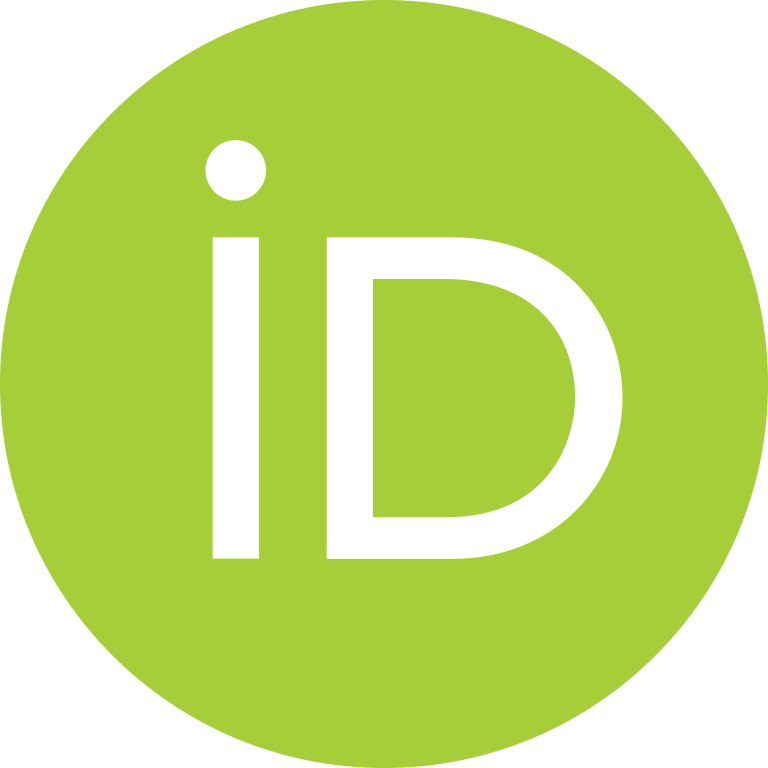}}}

\begin{document}

\title{An axially symmetric spacetime with causality violation in Ricci-inverse gravity}

\author{J. C. R. de  Souza \orcid{0000-0002-7684-9540} }
\email{jean.carlos@fisica.ufmt.br}
\affiliation{Instituto de F\'{\i}sica, Universidade Federal de Mato Grosso,\\
78060-900, Cuiab\'{a}, Mato Grosso, Brazil}

\author{A. F. Santos \orcid{0000-0002-2505-5273}}
\email{alesandroferreira@fisica.ufmt.br}
\affiliation{Instituto de F\'{\i}sica, Universidade Federal de Mato Grosso,\\
78060-900, Cuiab\'{a}, Mato Grosso, Brazil}

\begin{abstract}

In this paper, Ricci-inverse gravity is investigated. It is an alternative theory of gravity that introduces into the Einstein-Hilbert action an anti-curvature scalar that is obtained from the anti-curvature tensor which is the inverse of the Ricci tensor. An axially symmetric spacetime with causality violation is studied. Two classes of the model are discussed. Different sources of matter are considered. Then a direct relation between the content of matter and causality violation is shown. Our results confirm that Ricci-inverse gravity allows the existence of Closed Time-like Curves (CTCs) that lead to the violation of causality. Furthermore, a comparison is made between the results of general relativity and Ricci-inverse gravity. Other spacetimes, such as G\"{o}del and G\"{o}del-type universes, which are exact solutions of general relativity and allow for causality violations, are also explored in Ricci-inverse gravity framework.

\end{abstract}

\maketitle

\section{Introduction}

General relativity is an extraordinary gravitational model that describes physical reality very well and has undergone numerous tests since its formulation in 1915 \cite{review1, review2}. Although the theory and observational data confirm the success of general relativity, it is not a complete theory. There are two fundamental problems that require attention and solution. The classic version of gravity is complete, but a consistent quantum version has yet to be built. This fact shows that the gravitational interaction is different from other fundamental interactions described by the standard model of particle physics, since they have a well-known quantum version. Another open question in general relativity is: how to explain the recent accelerated expansion of the universe? The current accelerated cosmic expansion is a widely accepted fact in the scientific community and confirmed by several observational sources \cite{Riess,Per,Adam,Cole,Ande,expansao_ace,expansao_ace2}. In an attempt to find solutions to these problems, two different ways have been proposed in the literature: (i) introducing an exotic component of energy, called dark energy, into general relativity or (ii) modifying the Lagrangian of general relativity without resorting to dark energy. In this work, the second option is considered.

Alternative theories to general relativity have been a hot topic since Einstein proposed his model.  A historical review of the first attempts to extend general relativity, for different motivations, is presented in reference \cite{Hess}. For an overview of modified gravity theories, see references  \cite{Clifton, Shan}.  Here, Ricci-inverse gravity \cite{Amendola} is considered the alternative theory of gravity to be investigated. In this model, the modification consists of adding the anti-curvature tensor $A^{\mu\nu}$ in the Einstein-Hilbert action. The tensor $A^{\mu\nu}$ is defined as the inverse of the Ricci tensor $R_{\mu\nu}$. With this tensor the anti-curvature scalar $A=g_{\mu\nu}A^{\mu\nu}$ is defined.  It is important to emphasize that $A\neq R^{-1}$. In the reference \cite{Das} the Ricci-inverse theory is generalized and two classes of Ricci-inverse gravity are defined:  Class I and Class II. In Class I, the Lagrangian is proportional to a function $f(R, A)$ that depends on the Ricci scalar $R$ and the anti-curvature scalar $A$, while in Class II the function takes the form $f(R, A^{\mu\nu}A_{\mu\nu})$ which is a function of Ricci scalar and square of anti-curvature tensor. Ricci-inverse gravity has been investigated in several contexts, for example, anisotropic compact structures has been explored \cite{Sham},  the matter-antimatter asymmetry through baryogenesis in the realm of $f(R, A)$ theory has been analyzed \cite{Jaw}, a non-relativistic static and spherically symmetric cosmic structure embedded into a de Sitter cosmology has been investigated \cite{Scom} and no-go theorem for inflation in an extended Ricci-inverse gravity model has been studied \cite{Do1, Do}. Among the various topics investigated in this theory, an investigation into causality, its violation, and the existence of Closed Time-like Curves (CTCs) is missing. Here such a study is carried out.

To explore causality and the presence of CTCs in Ricci-inverse gravity, an axially symmetric metric is considered \cite{Haza, Ahmed, Ahmed1}. The main feature of this spacetime is the possibility of CTCs which are trajectories that allows objects to return to a point in their past. Another important point of this spacetime is that it satisfies the energy conditions. In such a study, different matter contents are analyzed. Furthermore, it is interesting to say that the existence of CTCs is not exclusive to axially symmetric metrics. Other widely studied solutions that lead to the violation of causality are the G\"{o}del metric \cite{K.Godel} and G\"{o}del-type metric \cite{Godel-type.Reboucas}, which are general relativity solutions with rotating matter and cosmological constant. These metrics are also analyzed in the context of Ricci-inverse gravity.

The present paper is organized as follows. In section II, an introduction to Ricci-inverse gravity is made. In section III, an axially symmetric metric in Ricci-inverse gravity is studied. The conditions for the existence of CTCs are presented. The set of field equations is solved in both general relativity and Ricci-inverse gravity. The results of the two theories are compared. It is shown that this alternative theory allows for the causality violation. Different matter contents are considered, such as a scalar field and an electromagnetic field. A discussion about G\"{o}del-type universes in Ricci-inverse gravity is made. In section IV, remarks and conclusions are presented.

\section{Ricci-inverse gravity}

In this section, a brief introduction to Ricci-inverse gravity is presented. This is a new way to modify the Einstein-Hilbert action proposed in \cite{Amendola}. This modification in Einstein's theory consists of the introduction of the anti-curvature tensor ($A^{\mu\nu}$) defined as  $A^{\mu \nu}R_{\nu \sigma} = \delta ^{\mu}_{\sigma}$. In the reference \cite{Das} two classes of Ricci-inverse gravity are introduced: Class I: the gravitational action is characterized by the function $f(R,A)$ which depends on Ricci ($R$) and anti-curvature ($A$) scalars. Class II: the theory is described by the function $f(R, A^{\mu\nu}A_{\mu\nu})$ which is a function of Ricci scalar and square of anti-curvature tensor. Here, Class I is considered. Then the action describing Ricci-inverse gravity is given as
\bea
S= \int d^4x \sqrt{-g}\left[f(R,A)-2\Lambda+{\cal L}_m\right].\label{0}
\eea
Taking  $f(R,A)=R+\kappa A$, Eq. (\ref{0}) becomes
\begin{equation}
    \label{1}
S= \int d^4x \sqrt{-g}\left[(R + \kappa A-2\Lambda)+{\cal L}_m\right],
\end{equation}
where $g$ is the metric determinant, $\kappa$ is the coupling constant, $R$ is the Ricci scalar, $A=g_{\mu\nu}A^{\mu\nu}$ is anti-curvature scalar, $\Lambda$ is the cosmological constant and ${\cal L}_m$ is the matter Lagrangian. This action can be written as
\bea
S= \int d^4x \sqrt{-g}\left[(g_{\mu \nu}R^{\mu \nu} + \kappa g_{\mu \nu} A^{\mu \nu}-2\Lambda)+{\cal L}_m\right].
\eea
Varying the action with respect to the metric leads to the field equations given by
\bea
&&R^{\mu \nu} - \frac{1}{2} R g^{\mu \nu}+\Lambda g^{\mu\nu} - \kappa A^{\mu \nu} - \frac{\kappa}{2}A g^{\mu \nu} + \frac{\kappa}{2}\Bigl [ 2g^{\rho \mu}\nabla _{\alpha } \nabla _{\rho } ( A^{\sigma \alpha} A^{\nu}_{ \sigma}) \nonumber\\
&&- \nabla ^2 (A^{\sigma \mu} A^{\nu}_{\sigma}) - g^{\mu \nu} \nabla _{\alpha} \nabla _{ \beta} ( A^{\sigma \alpha} A^{\beta}_{ \sigma}) \Bigl] = T^{\mu \nu},
\eea
where $\nabla_\mu$ denotes the covariant derivative and $T^{\mu\nu}$ is the energy-momentum tensor defined as
\begin{equation}\label{eq:2}
     T^{\mu\nu}=\frac{1}{\sqrt{-g}}\frac{\delta(\sqrt{-g}\mathcal{L}_{m})}{\delta g_{\mu\nu}}.
\end{equation}
Using that $A^{\alpha}_{\sigma} A^{\nu \sigma} = A^{\alpha \tau} g_{\tau \sigma} A^{\sigma \nu} = A^{\alpha \tau} A^{\nu}_{\tau} = A^{\alpha \sigma} A^{\nu}_{\sigma} = A^{\nu}_{ \sigma} A^{\alpha \sigma}$, the field equations become
\bea
\label{82}   
&&R^{\mu \nu} - \frac{1}{2} R g^{\mu \nu}+\Lambda g^{\mu\nu} - \kappa A^{\mu \nu} - \frac{\kappa}{2}A g^{\mu \nu} + \frac{\kappa}{2} \Bigl[ 2g^{\rho \mu}\nabla _{\alpha } \nabla _{\rho } ( A_{\sigma}^{ \alpha} A^{\nu \sigma})\nonumber\\
&& - \nabla ^2 (A^{\mu}_{\sigma} A^{\nu \sigma}) - g^{\mu \nu} \nabla _{\alpha} \nabla _{ \rho} ( A_{\sigma}^{ \alpha} A^{\rho \sigma}) \Bigl] = T^{\mu \nu}.
\eea
For simplicity, the Ricci-inverse gravity equations are written as
\bea
R^{\mu \nu} - \frac{1}{2} R g^{\mu \nu}+\Lambda g^{\mu\nu}+M^{\mu\nu}=T^{\mu \nu},\label{7}
\eea
where $M^{\mu\nu}$ contain the terms that modify the equations of general relativity due to the anti-curvature tensor and is defined as
\bea
M^{\mu\nu}= \kappa \left[A^{\mu \nu} + \frac{A}{2} g^{\mu \nu} - \frac{1}{2} \left( 2g^{\rho \mu}\nabla _{\alpha } \nabla _{\rho } ( A_{\sigma}^{ \alpha} A^{\nu \sigma})- \nabla ^2 (A^{\mu}_{\sigma} A^{\nu \sigma}) - g^{\mu \nu} \nabla _{\alpha} \nabla _{ \rho} ( A_{\sigma}^{ \alpha} A^{\rho \sigma}) \right)\right].
\eea

Our aim is to investigate cosmological solutions associated with causality violation. However, it is important to note that the Ricci-inverse gravity is constructed assuming that there is an anti-curvature tensor, i.e., $A^{\mu \nu}=R_{\mu \nu}^{-1}$.  Therefore, such a study is possible only in spacetime where there exists this quantity which is defined as
\begin{equation}
    \label{83}
    A^{\mu \nu}= \frac{1}{det[R_{\mu \nu}]}adj[R_{\mu \nu}].
\end{equation}
Note that, if the determinant of the Ricci tensor is null, there is no way to invert $R_{\mu \nu}$. Therefore, $det(R_{\mu \nu})\neq 0$ is a necessary condition to investigate a metric as a possible solution to Ricci-inverse gravity.

In the next section, an axially symmetric spacetime is studied in Ricci-inverse gravity. The main objective is to investigate whether this gravitational model allows solutions that leads to causality violation. In addition to the axially symmetric metric, other solutions of general relativity, such as G\"{o}del and G\"{o}del-type solutions,  are discussed in the approach described by Ricci-inverse gravity.

\section{Axially symmetric metric in Ricci-inverse gravity}

Here, an axially symmetric metric is investigated in Ricci-inverse gravity. The main feature of this solution is the presence of Closed Timelike Curves (CTCs), as discussed in references \cite{Haza, Ahmed, Ahmed1}. The line element that describes this spacetime at $(t, r, \phi, z)$ coordinates is given as
\begin{equation}
    \label{106}
    \mathrm{d}s^2 = \frac{\mathrm{d}r^2}{\alpha ^2 r^2} + r^2 \mathrm{d}z^2 +\left (-2r^2\mathrm{d}t + \frac{\beta z \mathrm{d}r}{r^2}-tr^2 \mathrm{d}\phi \right)\mathrm{d}\phi,
\end{equation}
where $\alpha$ and $\beta$ are non-zero constants, with $\beta>0$. It should be noted from Eq. (\ref{106}) that this spacetime has a coordinate singularity at $r = 0$.

The most notable characteristic of this metric is its ability to display CTCs. These curves can be obtained considering $r=r_0$, $z=z_0$ and $t=t_0$, with $r_0, z_0, t_0=\mathrm{const.}$ in Eq. (\ref{106}). Then
\bea
\mathrm{d}s^2 = -tr^2 \mathrm{d}\phi^2.\label{CTC}
\eea
This leads to three different curves: (i) null-like curve for $t=0$; (ii) space-like curve for $t_0<0$ and (iii) time-like curve for $t_0>0$. Therefore, in this spacetime, CTCs appear in an instant of time $t=t_0>0$.

In order to study this axially symmetric metric in Ricci-inverse gravity and compare this result with those obtained in general relativity, let's first review this solution in Einstein's theory. For this proposal, some geometric elements associated with the metric are necessary. Explicitly, the metric components are given as
\begin{eqnarray}
    \label{107}
    \nonumber g_{02}&=&g_{20}=-r^2, \\
    \nonumber g_{11}&=&\frac{1}{\alpha ^2 r^2}, \\
    \nonumber g_{12}&=&g_{21}=\frac{\beta z}{2 r^2}, \\
    \nonumber g_{22}&=&-r^2 t, \\
     g_{33}&=& r^2,
\end{eqnarray}
and their inverses are
\begin{eqnarray}
    \label{108}
    \nonumber g^{00}&=&\frac{\alpha ^2 \beta ^2 z^2}{4 r^6}+\frac{t}{r^2}, \\
    \nonumber g^{01}&=&g^{10}=\frac{\alpha ^2 \beta z}{2 r^2}, \\
    \nonumber g^{02}&=&g^{20}=-\frac{1}{r^2}, \\
    \nonumber g^{11}&=&\alpha ^2 r^2, \\
     g^{33}&=& \frac{1}{r^2}.
\end{eqnarray}
These metric components lead to the non-zero Christoffel symbols 
\begin{eqnarray}
    \label{109}
    \nonumber \Gamma^{0}_{01} &=& \frac{1}{r},\quad\quad \Gamma^{0}_{02} = \frac{\alpha ^2 \beta z+r}{2 r},\quad\quad\Gamma^{0}_{11} = \frac{\beta z}{2 r^5}, \\
    \nonumber \Gamma^{0}_{12} &=&  -\frac{\alpha ^2 \beta ^2 z^2}{4 r^5},\quad\quad \Gamma^{0}_{13} =  -\frac{\beta}{4 r^4},\\
    \nonumber \Gamma^{0}_{22} &=& \frac{\alpha ^2 \beta ^2 z^2+4 \alpha ^2 \beta r^3 t z+4 r^4 t}{8 r^4},\quad\quad \Gamma^{0}_{23} = \frac{\alpha ^2 \beta ^2 z}{8 r^4},\\
    \nonumber \Gamma^{0}_{33} &=& -\frac{\alpha ^2 \beta z}{2 r},\quad\quad \Gamma^{1}_{02} = \alpha ^2 r^3,\quad\quad \Gamma^{1}_{11} = -\frac{1}{r},  \\
    \nonumber \Gamma^{1}_{12} &=& -\frac{\alpha ^2 \beta z}{2 r},\quad\quad \Gamma^{1}_{22} = \frac{1}{4} \alpha ^2 \left(\beta z+4 r^3 t\right),\\
    \nonumber \Gamma^{1}_{23} &=& \frac{\alpha ^2 \beta}{4},\quad\quad \Gamma^{1}_{33} = -\alpha ^2 r^3,\quad\quad \Gamma^{2}_{12} =  \frac{1}{r},\\
    \nonumber \Gamma^{2}_{22} &=& -\frac{1}{2},\quad\quad \Gamma^{3}_{12} =  -\frac{\beta}{4 r^4},\quad\quad \Gamma^{3}_{13} =\frac{1}{r}. 
\end{eqnarray}
Thus, the non-zero components of the Ricci tensor are
\begin{eqnarray}
    \label{110}
    \nonumber R_{02} &=& 3 \alpha ^2 r^2,\\
    \nonumber R_{11} &=& -\frac{3}{r^2},\\
    \nonumber R_{12} &=& -\frac{3 \alpha ^2 \beta z}{2 r^2},\\
    \nonumber R_{22} &=& \frac{\alpha ^2 \left(\beta ^2+24 r^6 t\right)}{8 r^4},\\
    R_{33} &=& -3 \alpha ^2 r^2.
\end{eqnarray}
and in the contravariant form become
\begin{eqnarray}
    \label{111}
    \nonumber R^{00} &=& \frac{\alpha ^2 \left[\beta ^2 \left(1-6 \alpha ^2 r^2 z^2\right)-24 r^6 t\right]}{8 r^8}, \\
    \nonumber R^{01} &=& -\frac{3 \alpha ^4 \beta z}{2 r^2},\\
    \nonumber R^{02} &=& \frac{3 \alpha ^2}{r^2},\\
    \nonumber R^{11} &=& -3 \alpha ^4 r^2\\
    R^{33} &=& -\frac{3 \alpha ^2}{r^2}.
\end{eqnarray}
The Ricci scalar is given as
\bea
\nonumber R=g_{\mu \nu}R^{\mu \nu}&=&-12 \alpha ^2.
\eea

With these ingredients, the field equations of general relativity,
\begin{equation}
    \label{117}
    R^{\mu \nu}-\frac{1}{2}Rg^{\mu \nu}+\Lambda g^{\mu \nu}=T^{\mu \nu},
\end{equation}
can be solved. For such an analysis, pure radiation is chosen as the matter content whose energy-momentum tensor is defined as
\begin{equation}
    \label{116}
    T^{\mu \nu}=\rho \zeta ^{\mu} \zeta ^{\nu},
\end{equation}
with $\zeta ^{\mu} = (1,0,0,0)$ being a null vector. Then the field equations of general relativity are given as
\begin{eqnarray}
    \label{118}
 \rho &=& \frac{6 \alpha ^4 \beta ^2 r^2 z^2+\alpha ^2 \left[\beta ^2 \left(2 \Lambda  r^2 z^2+1\right)+24 r^6 t\right]+8 \Lambda  r^6 t}{8 r^8}, \\
0&=& \frac{\alpha ^2 \beta z \left(3 \alpha ^2+\Lambda \right)}{2 r^2}, \\
0&=&-\frac{3 \alpha ^2+\Lambda }{r^2},\\
0&=&\alpha ^2 r^2 \left(3 \alpha ^2+\Lambda \right), \\
0&=&\frac{3 \alpha ^2+\Lambda }{r^2}.
\end{eqnarray}
This set of equations leads to the solution
\begin{eqnarray}
    \label{119}
    \nonumber \rho &=& \frac{\alpha ^2 \beta ^2}{8 r^8},\\
    \Lambda &=&-3 \alpha ^2.
\end{eqnarray}
Therefore, the axially symmetric metric is a solution of general relativity with a negative cosmological constant and an energy density that decreases with $r$  \cite{Haza, Ahmed}.

Now the goal is to study this cosmological solution in Ricci-inverse gravity and discuss whether CTCs are present in this gravitational theory. To calculate the anti-curvature tensor using Eq. (\ref{83}), it is first necessary to analyze the determinant of the Ricci tensor. For the axially symmetric metric one finds that
\begin{equation}
    \label{9}
    det[R_{\mu \nu}] = -81 \alpha ^6 r^4.
\end{equation}
Then, as the determinant of the Ricci tensor is not zero,  it is possible to determine all components of the anti-curvature tensor. The non-zero components are
\begin{eqnarray}
    \label{112}
    \nonumber A^{00} &=&-\frac{6 \alpha ^2 \beta ^2 r^2 z^2+\beta ^2+24 r^6 t}{72 \alpha ^2 r^8},\\
    \nonumber A^{01} &=&-\frac{\beta z}{6 r^2},\\
    \nonumber A^{02} &=& \frac{1}{3 \alpha ^2 r^2} ,\\
    \nonumber A^{11} &=& -\frac{r^2}{3},\\
    A^{33} &=& -\frac{1}{3 \alpha ^2 r^2},
\end{eqnarray}
and in the covariant form become
\begin{eqnarray}
    \label{113}
    \nonumber A_{02} &=&\frac{r^2}{3 \alpha ^2}, \\
    \nonumber A_{11} &=& -\frac{1}{3 \alpha ^4 r^2},\\
    \nonumber A_{12} &=& -\frac{\beta z}{6 \alpha ^2 r^2},\\
    \nonumber A_{22} &=& -\frac{\beta ^2-24 r^6 t}{72 \alpha ^2 r^4},\\
    A_{33} &=& -\frac{r^2}{3 \alpha^2}.
\end{eqnarray}
And the anti-curvature scalar is given as
\bea
 A=g_{\mu \nu}A^{\mu \nu}&=&-\frac{4}{3 \alpha ^2}.
\eea
With these quantities and considering pure radiation given in the energy-momentum tensor (\ref{116}) as the source of matter, the Ricci-inverse gravity field equations, Eq. (\ref{7}), become
\begin{eqnarray}
 \rho &=&\frac{1}{216 \alpha ^2 r^8} \left\{162 \alpha ^6 \beta ^2 r^2 z^2+27 \alpha ^4 \left[\beta ^2 \left(2 \Lambda  r^2 z^2+1\right)+24 r^6 t\right]+\right\}\\
\nonumber &&+\frac{1}{216 \alpha ^2 r^8}\left\{54 \alpha ^2 \left(\beta ^2 \kappa r^2 z^2+4 \Lambda  r^6 t\right)- 35 b^2 \kappa+216 \kappa r^6 t \right\},\\
 0 &=& \frac{\beta z \left(3 \alpha ^4+\alpha ^2 \Lambda +\kappa \right)}{2 r^2},\\
 0&=&-\frac{3 \alpha ^4+\alpha ^2 \Lambda +\kappa}{\alpha ^2 r^2},\\
0&=&r^2 \left(3 \alpha ^4+\alpha ^2 \Lambda +\kappa \right),\\
0&=&\frac{3 \alpha ^4+\alpha ^2 \Lambda +\kappa}{\alpha ^2 r^2}.
\end{eqnarray}
Solving this set of equations for the energy density $\rho$ and the cosmological constant $\Lambda$, we get
\bea
   \nonumber \rho &=& \frac{\beta ^2 \left(27 \alpha ^4-35 \kappa \right)}{216 \alpha ^2 r^8},\\
    \Lambda &=& -\frac{3 \alpha^4+\kappa}{\alpha ^2}.    \label{128}
\eea
With the above results, it can be stated that the axially symmetric metric is a solution to the Ricci-inverse gravity field equations. Therefore, this gravitational theory presents CTCs as defined in Eq. (\ref{CTC}), and as a consequence, causality is violated. Comparing with the results of general relativity, Eq. (\ref{119}), it is noted that in both cases the cosmological constant assumes a negative value. With regard to energy density, some observations must be made. In general relativity, the energy density satisfies the energy condition and is always positive  \cite{Haza, Ahmed}. In Ricci-inverse gravity, to obtain a positive energy density, a relation arises between the constant $\alpha$ and the coupling constant $\kappa$, i.e., $\kappa<\frac{27\alpha^4}{35}$. Also, it is important to note that the results of general relativity are recovered when the coupling constant $\kappa$ becomes zero.

In addition to these results that lead to a non-causal universe in Ricci-inverse gravity, other analyzes have been developed considering different sources of matter. Two matter contents have been chosen: (i) a scalar field whose energy-momentum tensor is given as
\begin{equation}
      \label{129}
    T^{\mu \nu (S)}= \partial ^{\mu} \phi \partial ^{\nu} \phi - \frac{1}{2}g^{\mu \nu} g_{\rho \lambda}\partial ^{\rho} \phi \partial ^{\lambda} \phi,
\end{equation}
and (ii) an electromagnetic field described by energy-momentum tensor
\begin{equation}
    \label{130}
    T^{\mu \nu(EM)}=-F^{\mu \alpha}F^{\nu}_{\alpha}+\frac{1}{4}g^{\mu \nu}F_{\beta \alpha}F^{\alpha \beta}.
\end{equation}
Using Eqs. (\ref{129}) and (\ref{130}) in Eq. (\ref{7}), two sets of field equations are obtained. However, there are no solutions to these sets of equations. Therefore, these matter sources prevent the axially symmetric metric given in Eq. (\ref{106}) from being a solution of Ricci-inverse gravity. This implies that the causality violation generated by this metric is avoided for proper matter content.

Although the main objective has been to study the axially symmetric metric with causality violation, other metrics that present CTCs have also been investigated in the context of Ricci-inverse gravity. The G\"{o}del metric \cite{K.Godel} is an exact solution of general relativity, but it is not a solution in Ricci-inverse gravity. The same result occurs for the G\"{o}del-type metric \cite{Godel-type.Reboucas}. In fact, these metrics have that the determinant of the Ricci tensor is zero, i.e., the inverse of the Ricci tensor is not determined. This means that the anti-curvature tensor for both metrics can not be calculated.  Therefore, these metrics fail the first test to be studied in  Ricci-inverse gravity.

\section{Conclusion}

An alternative theory of gravity has been considered. An anti-curvature tensor that is defined as the inverse of the Ricci tensor is used to construct the Ricci-inverse gravity. In this theory an axially symmetric spacetime is studied. This metric allows the existence of CTCs, closed curves in time that permit travel to the past. As a consequence, the violation of causality arises. Considering pure radiation as matter content, our results show that the axially symmetric metric is a solution in Ricci-inverse gravity. Then this gravitational model allows for the violation of causality. It is found that the cosmological constant has a negative value as obtained in general relativity. But, a positive energy density, which satisfies the energy conditions, requires a condition involving the Ricci-inverse gravity coupling constant and the constant $\alpha$, a free parameter of the metric. These results are consistent with the class I of the model which is defined as a function of the form $f(R,A)$. Class II, which is characterized by a function of Ricci scalar and square of the anti-curvature tensor, is also analyzed. But the axially symmetric metric is not a solution for this set of equations. For different sources of matter, such as scalar field and electromagnetic field, it is shown that this spacetime is not a solution for the gravitational model. Therefore, our results indicate that there is a relation between the matter content of the universe and the existence of CTCs or causality violation. Furthermore, the G\"{o}del and G\"{o}del-type universes were also investigated in Ricci-inverse gravity. Our analysis shows that these metrics lead to a $det(R_{\mu\nu})=0$, then there is no Ricci tensor inverse. Thus, it is impossible to construct an anti-curvature tensor associated with the G\"{o}del-types universes. Therefore, these exact solutions of general relativity do not satisfy the mandatory condition to be investigated in Ricci-inverse gravity. Here a natural question arises: how to investigate metrics that display $det(R_{\mu\nu})=0$ in Ricci-inverse gravity? Are they simply excluded? or are they a problem of this alternative theory of gravity? These are open questions and are under investigation.

\section*{Acknowledgments}

This work by A. F. S. is partially supported by National Council for Scientific and Technological Develo\-pment - CNPq project No. 313400/2020-2. J. C. R. S. thanks CAPES for financial support.


\global\long\def\link#1#2{\href{http://eudml.org/#1}{#2}}
 \global\long\def\doi#1#2{\href{http://dx.doi.org/#1}{#2}}
 \global\long\def\arXiv#1#2{\href{http://arxiv.org/abs/#1}{arXiv:#1 [#2]}}
 \global\long\def\arXivOld#1{\href{http://arxiv.org/abs/#1}{arXiv:#1}}


\end{document}